\documentclass[12pt,twocolumn]{aastex631}
\usepackage{color}
\usepackage{hyperref}
\usepackage{epstopdf}
\usepackage{graphicx}
\usepackage[FIGTOPCAP]{subfigure}
\usepackage{CJKutf8}
\usepackage{amsmath}
\usepackage{longtable}
\begin{document}
	\title{GRB 201104A: A “Repetitive” Short Gamma-Ray Burst?}
	
	\correspondingauthor{Yun Wang}
	\email{wangyun@pmo.ac.cn}
	
	\author[0000-0002-8385-7848]{Yun Wang}
	\affiliation{Key Laboratory of Dark Matter and Space Astronomy, Purple Mountain Observatory, Chinese Academy of Sciences, Nanjing 210034, China}
	\affiliation{School of Astronomy and Space Science, University of Science and Technology of China, Hefei, Anhui 230026, China}
	
	
	\author{Lu-Yao Jiang}
	\affiliation{Key Laboratory of Dark Matter and Space Astronomy, Purple Mountain Observatory, Chinese Academy of Sciences, Nanjing 210034, China}
	\affiliation{School of Astronomy and Space Science, University of Science and Technology of China, Hefei, Anhui 230026, China}
	
	\author[0000-0002-9037-8642]{Jia Ren}
	\affiliation{School of Astronomy and Space Science, Nanjing University, Nanjing 210093, China}
	\affiliation{Key Laboratory of Modern Astronomy and Astrophysics (Nanjing University), Ministry of Education, China}
	
	
	\begin{abstract}
		Gamma-ray bursts (GRBs) are divided into short gamma-ray bursts (SGRBs) and long gamma-ray bursts (LGRBs) based on the bimodal distribution of their durations. 
		LGRBs and SGRBs are typically characterized by different statistical characteristics.
		Nevertheless, there are some samples that challenge such a framework, such as GRB 060614, a long-duration burst with short-burst characteristics.
		Furthermore, GRBs are generally considered to be an event with no periodic or repetitive behavior, since the progenitors usually undergo destructive events,
		such as massive explosions or binary compact star mergers.
		In this work, we investigated Fermi data for possible quasiperiodic oscillations and repetitive behaviors of GRBs using timing analysis methods and report a special event GRB 201104A,
		which is a long-duration burst with the characteristics of an SGRB, and it exhibits a ``repetitive" behavior.
		We propose that such a situation may arise from lensed SGRBs and attempt to verify it by Bayesian inference.
		In addition, we extend the spectral analysis to Bayesian inference.
		In spite of the existence of at least two distinct time periods with a nearly identical spectrum,
		there is no strong evidence that they result from a lensing GRB.	
		Taking the gravitational-lensing scenario out of consideration, a long burst would resemble a short burst in its repetitive behavior,
		which presents a challenge for the current classification scheme.	
		
	\end{abstract}
	\keywords{Gamma-ray burst; Gravitational lensing}
	
	\section{Introduction} \label{sec:intro}
	The gamma-ray bursts (GRBs) is the brightest explosion in the universe and is characterized by high variability.
	GRBs can be divided into short gamma-ray bursts (SGRBs) and long gamma-ray bursts (LGRBs) based on the bimodal distribution of the observed durations in the BATSE era \citep{kouveliotou1993identification}.
	LGRBs originate in star-forming regions in galaxies and are observed in association with massive star-collapse supernovae \citep{woosley1993gamma,fruchter2006long}.
	Short bursts are located in low star-forming regions of their host galaxies, and they are believed to originate from compact binaries {\citep{eichler1989nucleosynthesis,narayan1992gamma,gehrels2005short,fong2009hubble,leibler2010stellar,fong2013locations,berger2014short,tunnicliffe2014nature,o2022deep}.}
	The duration of the burst ($T_{90}$) is often inconsistent across energy ranges and different instruments {\citep{qin2012comprehensive,bromberg2013short}}, so other characteristics of GRBs are important to consider.
	{For example,}
	the delay in the arrival of photons of different energies \citep{norris2000connection,norris2006short},
	the $E_{\gamma, {\rm iso}}$ and $E_{p,z}$ correlation \citep{amati2002intrinsic}, as well as the star formation rate of the host galaxy \citep{li2016comparative}.
	{ in the spectral analysis results of GRBs observed by Fermi, the low-energy spectral index of long bursts is lower than that of short bursts \citep{nava2011spectral}.}
	However, there are still some special cases, such as long-duration bursts with short-burst characteristics {(GRB 060614 \citep{gehrels2006new,yang2015possible,jin2015light} and GRB 211211A \citep{rastinejad2022kilonova,gompertz2022minute,yang2022peculiarly})} and short-duration bursts with long burst characteristics {(GRB 040924 \citep{fan2005optical} and GRB 200826A \citep{zhang2021peculiarly,ahumada2021discovery,2022ApJ...932....1R})}.
	
	All these issues are intertwined, so it is challenging to interpret all observed properties in a consistent picture.
	The confusion stems from our lack of understanding of the central engine and dissipation mechanisms.
	Studying the power density spectra of the prompt emission of GRBs can help to solve these puzzles \citep{beloborodov1998self,dichiara2013average}.
	While most observations and theories indicate that GRBs do not repeat \citep{meszaros2006gamma,kumar2015physics},
	some work has attempted to explore the possibility of quasi-periodic oscillation (QPO) signals in GRBs \citep{dichiara2013search,zhang2016central,tarnopolski2021comprehensive}.
	{However, some bursts with repetitive behavior on the light curve are thought to be caused by gravitational lensing \citep{paynter2021evidence,wang2021grb,veres2021fermi,lin2021search}.}
	In this work, we investigated possible QPOs and repetitive behaviors in the prompt emission of GRBs observed by Fermi.
	We use the Bayesian test to find and confirm periodic or quasiperiodic oscillations in red noise, which builds upon the procedure {outlined by} \cite{vaughan2010bayesian}.
	To identify possible repetitive events, we calculate the autocorrelation function proposed by \cite{paynter2021evidence}.
	We have found a {special example}, GRB 201104A, and we have taken into consideration the possible repetition artifacts caused by the lens effect.
	Furthermore, we extend the spectral analysis to Bayesian inference of lens and nonlens models.
	We believe that this procedure should be taken into account in future research to certify gravitationally lensed GRBs.	
	
	The paper is organized as follows:
	In Section \ref{sec:search_method}, we describe the search method we use to identify repeating events and QPOs in Fermi's GRBs.
	In Section \ref{sec:obs_ana}, we present observations of GRB 201104A and a detailed analysis of its properties.
	In Section \ref{sec:lens}, we performed Bayesian inferences under the lensing hypothesis, considering both light curves and spectrum data.
	In Section \ref{sec:sd}, we summarize our results with some discussion.
	
	\section{Comprehensive search} \label{sec:search_method}
	In modern astrophysics, the analysis of time series is an essential tool.
	These two methods were used to search for possible QPO and repetitive behaviors in the current Fermi-GBM data \citep{meegan2009fermi}.
	
	\subsection{Autocorrelation function} \label{sec:ACF}
	Signal autocorrelation can be used to measure the time delay between two temporally overlapping signals.
	It may be an {intrinsic property of the source}, or it may be due to gravitational lensing.
	The standard autocorrelation function (ACF) is defined as follows:
	\begin{equation}
		C(k) = \frac{\sum_{t=0}^{N-k} (I_t-\overline I)(I_{t+k}-\overline I)}
		{\sum_{t=0}^{N} (I_t-\overline I)^2}.
		\label{eq:correlation}
	\end{equation}
	To fit the ACF sequence, we apply the Savitzky-Golay filter $F(\delta t)$. 
	The values of the window length and the order of the polynomial are set to be 101 and 3, respectively \citep{paynter2021evidence}.
	The dispersion ($\sigma$) between the ACF and the fit $F(k)$ is
	\begin{equation}
		\sigma^2 = \frac{1}{N}\sum_{j=0}^{N} [C(k) - F(k)]^2,
		\label{eq:dispersion}
	\end{equation}
	where $N$ is the total number of bins.
	As usual, we identify the $3\sigma$ outliers as our candidates.
	
	\subsection{Power Density Spectra} \label{sec:PDS}
	To identify possible quasiperiods in the time series data, red noise must be modeled in order to assess the significance of any possible period.
	For the above purpose, we developed a procedure based on \cite{vaughan2005simple,vaughan2010bayesian,vaughan2013random} and \cite{covino2019gamma}.
	Also, we refer to \cite{beloborodov1998self,guidorzi2012average,guidorzi2016individual,dichiara2013average,dichiara2013search} for details on analyzing power density spectra in GRBs.
	Power density spectra (PDS) are derived by discrete Fourier transformation and normalized according to \cite{leahy1983searches}.
	
	We consider two models to fit the PDS \citep{guidorzi2016individual}, the first is a single power-law function plus white noise called a power-law model,
	\begin{equation}
		S_{\rm PL}(f) = N\,f^{-\alpha} + B
		\qquad ,
		\label{}
	\end{equation}
	here $N$ is a normalization factor, $f$ is the sampling frequency, and its lower limit is related to the length of the time series, which is $1/T$ (the time interval).
	The upper limit of $f$ is the Nyquist frequency, which is $1/(2{\delta_t})$, where $\delta_t$ is the time bin size of data.
	The value of white noise $B$ is expected to be two, which is the expected value of a $\chi{^2_2}$ distribution for pure Poissonian variance in the Leahy normalization.
	In some GRBs, there is a distinct broken power law. As such, we consider another model called BPL,
	\begin{equation}
		S_{\rm BPL}(f) = N\,\Big[1 + \Big(\frac{f}{f_{\rm b}}\Big)^{\alpha}\Big]^{-1} + B
		\qquad ,
		\label{}
	\end{equation}
	There is one more parameter here, the break frequency $f_b$.
	We employ a Bayesian inference \citep{thrane2019introduction,van2021bayesian} approach for parameter estimation and model selection by using the nested sampling algorithm Dynesty \citep{speagle2020dynesty,skilling2006nested,higson2019dynamic} in {\tt Bilby} \citep{ashton2019bilby}.
	The maximum likelihood function we use is called $Whittle$ likelihood function \citep{vaughan2010bayesian}.
	After we have the posterior distribution of the model parameters, we calculate the global significance of every frequency in the PDS according to $T_\text{R} = {\rm max_j}
	R_j $, where $R = 2P/S$, $P$ is the simulated or observed PDS, and $S$ is the best-fit PDS model. 
	This method selects the maximum deviation from the continuum for each simulated PDS.
	The observed $T_\text{R}$ values are compared to the simulated distribution and significance is assessed directly.
	The corrections for the multiple trials performed were included in the analysis because the same procedure was applied to the simulated as well as to the real data.
	
	\section{Observation and Data analysis}\label{sec:obs_ana}
	We analyzed the GRBs in the Fermi GBM Burst Catalog \citep{gruber2014fermi,bhat2016third,von2014second,von2020fourth} using the above method.
	In total, we examined the PDS of 248 short bursts and 920 long bursts (bounded by $T_{90}$ of 3 seconds) with peak counts greater than 50 in 64 ms {time} bin.
	Among these samples, we did not find a quasi-periodic signal with a significance exceeding 3 $\sigma$, but among the candidates of the ACF check, we found an {interesting example},which is GRB 201104A.
	The basic observation information and data reduction are as follows.
	
	The Fermi-GBM team reports the detection of a possible long burst GRB 201104A (trigger 626140861.749996 / 201104001) \citep{2020GCN.28823....1F}.
	At the same time, Fermi-LAT \citep{atwood2009large} also observed high-energy high photons from this source with high significance \citep{2020GCN.28828....1O}.
	In addition, we try to search the observation data of other telescopes through GCN \footnote{https://gcn.gsfc.nasa.gov},
	but there is no X-ray and optical data available with other telescopes.
	
\begin{figure}
		\centering
		\includegraphics[width=0.4\textwidth]{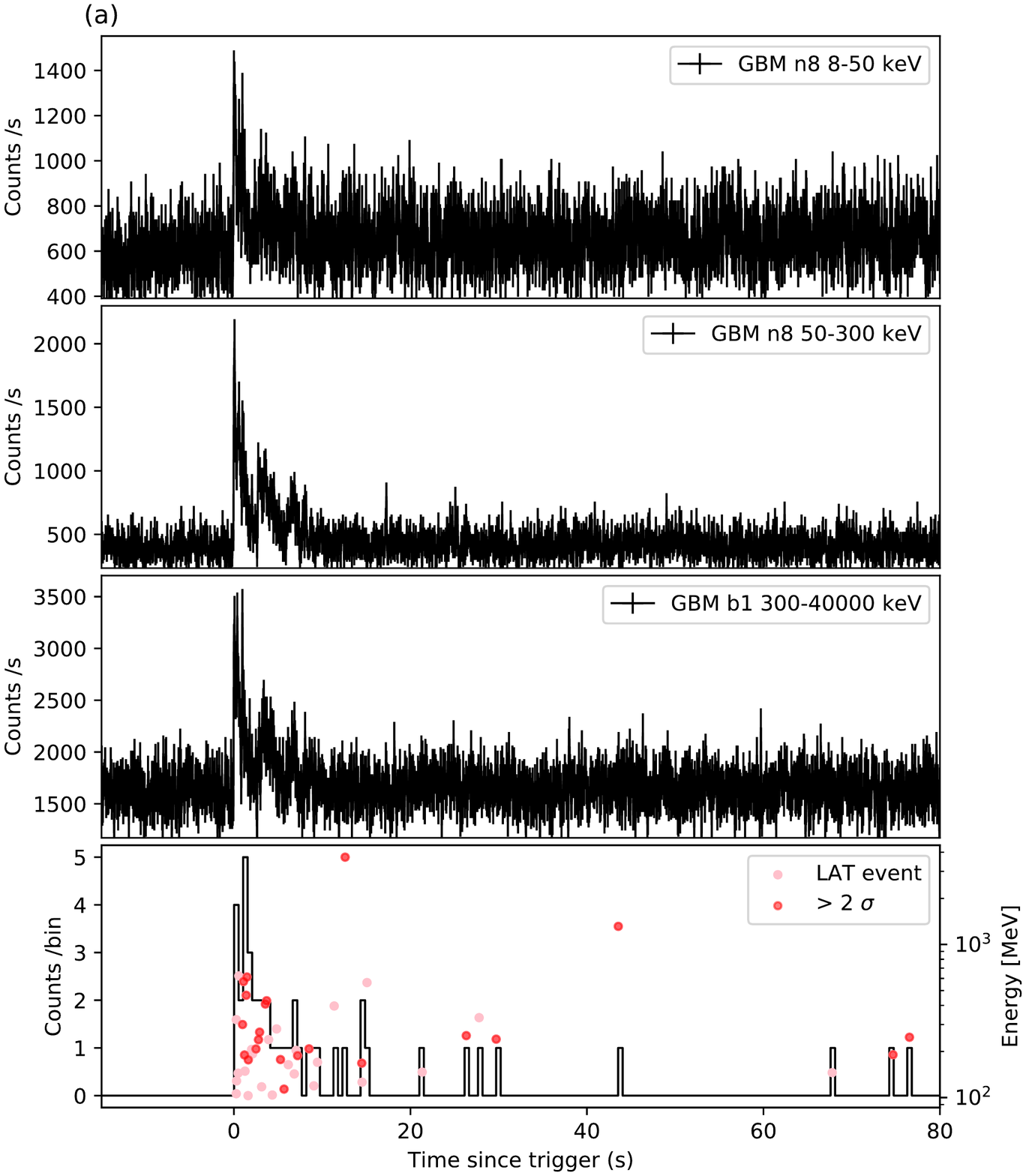}
		\includegraphics[width=0.4\textwidth]{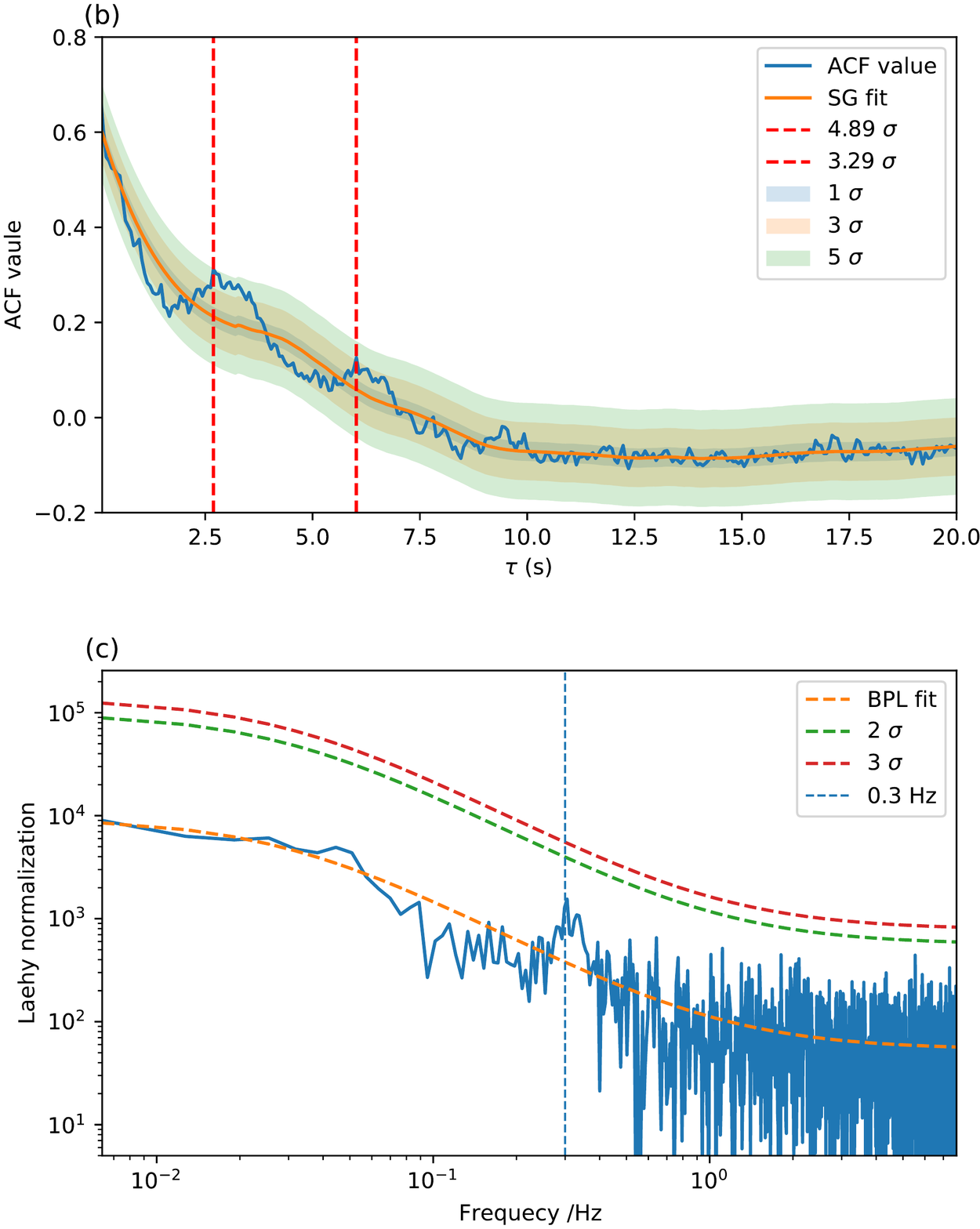}
		\caption{Observational and time series analysis of GRB 201104A.
			(a) is the light curve of each energy band of different instruments.
			The binsizes of the GBM and LAT are set to 64 ms and 512 ms, respectively.
			The pink dots in the bottommost panel represent each LAT event,
			while events with confidence exceeding 2 $\sigma$ are filled in red.
			(b) is the result of the autocorrelation of the light curve of the NaI detector,
			the energy range is 50-300 keV, and the binsize is 64 ms.
			The solid orange lines indicate the fits to the light curves by using a Savitzky-Golay smoothing filter with a 101 window length.
			The shaded regions in colors show the 1$\sigma$, 3$\sigma$, and 5$\sigma$ containment bands of the Savitzky–Golay fit.
			The maximum significance and the corresponding time are also indicated in the plot.
			The red dashed line indicates the time corresponding to exceeding 3$\sigma$ significance.
			(c) is the power density spectra of the same light curve mentioned in (b);
			the best-fit model (BPL model) and the significance at each frequency are shown as different colored dashed lines.
			The power density spectrum exhibits a peak at 0.3 Hz.}
		\label{fig:1}
	\end{figure}

	We present a further analysis of GRB 201104A observed by Fermi instruments in this work.
	Fermi-GBM  \citep{meegan2009fermi} has 12 sodium iodide (NaI) detectors and 2 bismuth germanate (BGO) detectors.
	According to the pointing direction and count rate of the detectors, we selected a NaI detector (n8) and a BGO detector (b1) respectively.
	The Fermi-GBM data is processed with {\tt GBM Data Tools}  \citep{GbmDataTools}, which makes it incredibly easy for users to customize.
	We performed a standard unbinned maximum likelihood analysis for GRB \footnote{https://fermi.gsfc.nasa.gov/ssc/data/analysis/scitools} using {\tt Fermitools} \footnote{https://github.com/fermi-lat/Fermitools-conda/wiki} (version 2.0.8),
	and we determined the probability of each photon originating from this source.
	In Figure \ref{fig:1} (a), we present the GBM and LAT light curves for several energy bands.
	It appears to be a long burst of three episodes and with a weak extending component.
	ACF and PDS results are also shown in (b) and (c) in Figure \ref{fig:1}.
	A double peak in the ACF means that the light curve is similar after two time shifts, while there is also a peak in the PSD at a frequency of 0.3 Hz.
	While the significance level of this quasi-periodicity is not high enough, we also analyze in detail each possible episode in the following section.
	
	\subsection{Spectral Analysis}\label{sec:spec_ana}
\begin{figure}
		\centering
		\includegraphics[width=0.4\textwidth]{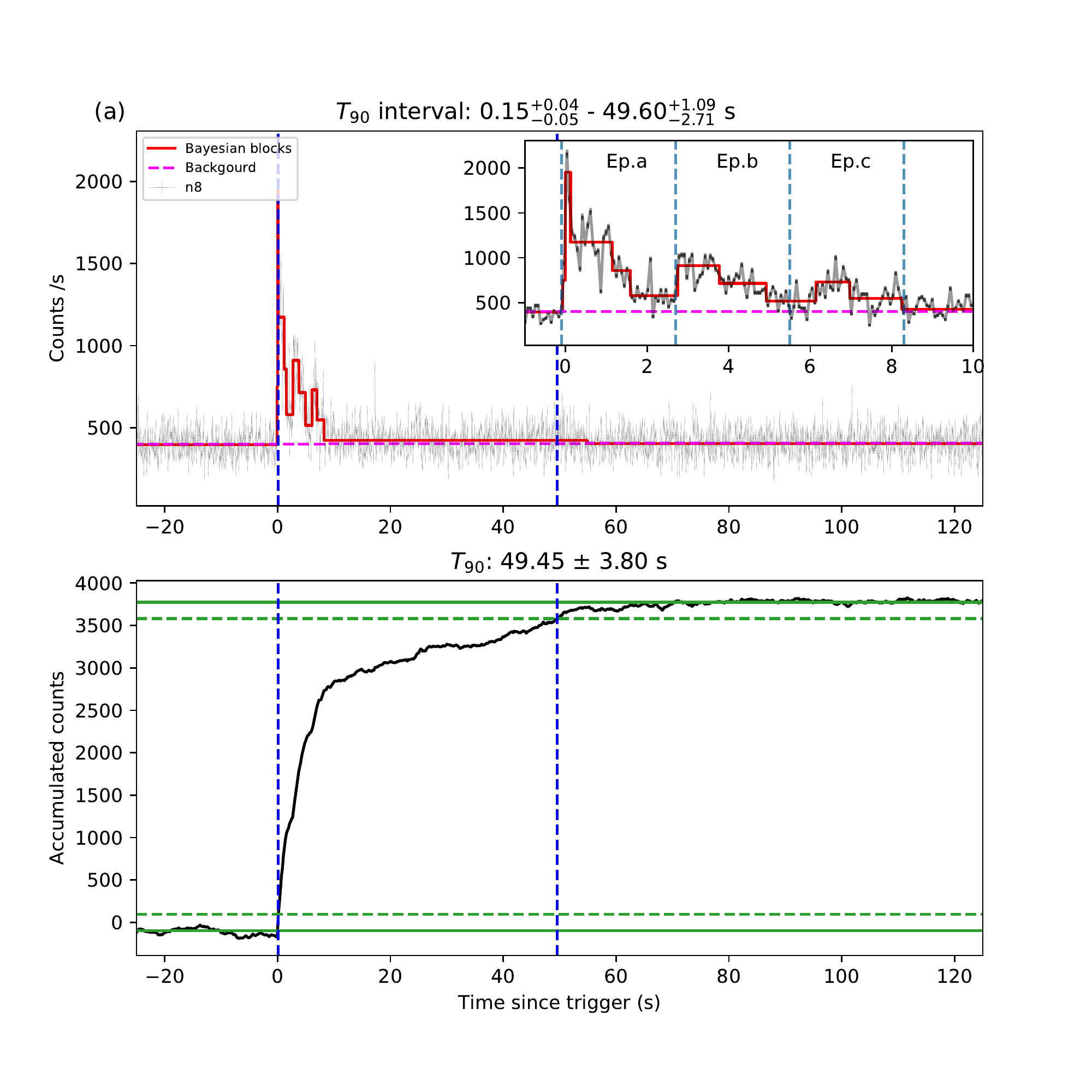}
		\includegraphics[width=0.4\textwidth]{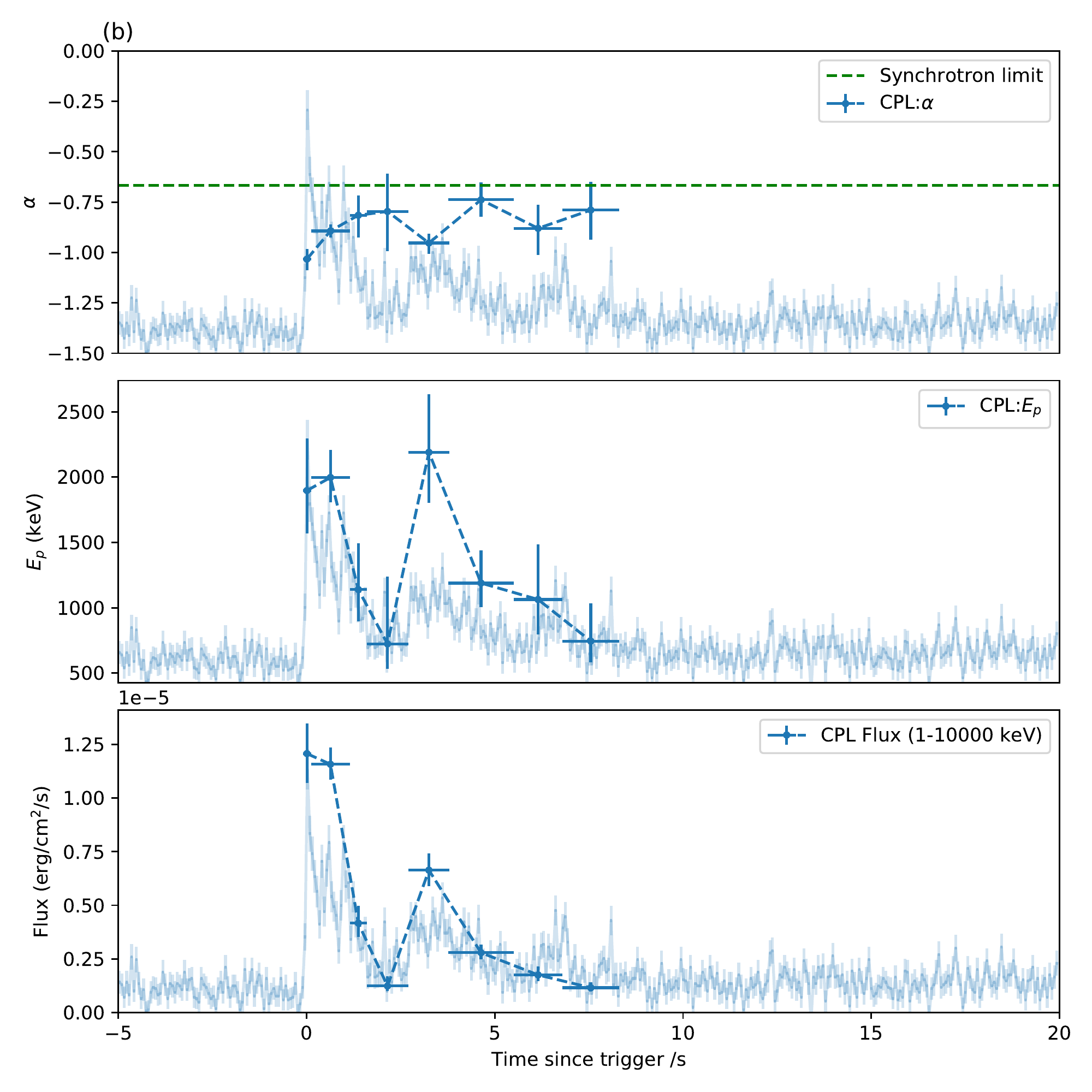}
		\caption{Further analysis based on observational data.
			(a) is the $T_{90}$ calculation. The upper panel of (a) show the light curve of the NaI detector,
			the energy range is 50-300 keV, and the binsize is 64 ms.
			The red solid line and the magenta dashed line represent the Bayesian block and the background, respectively.
			And the subplots show the time interval for each epsiode.
			The bottom {panel} of (a) show the photon count accumulation curve.
			The green dashed line represents the interval from 5$\%$ to 95$\%$ of the cumulative photon count.
			(b) is the temporal evolution of spectral parameters $\alpha$, $E_p$ and Flux of CPL model. 
			In the upper panel of (b), the green dashed line represents the limit of the synchrotron shock model,
			also known as the ``Line of Death" \citep{preece1998synchrotron}.}
		\label{fig:2}
	\end{figure}

	In order to confirm the duration of this burst, we recalculated the $T_{90}$ \citep{koshut1996systematic} of the GBM n8 detector in the energy range of 50-300 keV.
	We then used the Bayesian block technique \citep{scargle2013studies} to determine the time interval of this burst (see the left of Figure \ref{fig:2}).
	We divide the main burst [$T_0$ -0.1 , $T_0$ + 8.3 s] into three episodes,
	which are Episode a [$T_0$ -0.1 , $T_0$ + 2.7 s], Episode b [$T_0$ + 2.7, $T_0$+ 5.5 s], and Episode c [$T_0$ + 5.5, $T_0$ + 8.3 s].
	We perform both time-integrated and time-resolved spectral analyses of GRB 201104A, and the specific time interval is shown in Table \ref{tab:tab1}.
	For each time interval, we extract the corresponding source spectra,
	background spectra and instrumental response files following the procedure described Section \ref{sec:specinf} and \cite{GbmDataTools}. 
	In general, the energy spectrum of GRB can be fitted by an empirical smoothly joined broken power-law function (the so-called “Band” function; {\citealt{band1993batse}}).
	The Band function takes the form of 
	\begin{footnotesize}
	\begin{equation}
		N(E)=
		\begin{cases}
			A(\frac{E}{100\,{\rm keV}})^{\alpha}{\rm exp}{(-\frac{E}{E_0})}, \mbox{if $E<(\alpha-\beta)E_{0}$ }\\
			A[\frac{(\alpha-\beta)E_0}{100\,{\rm keV}}]^{(\alpha-\beta)}{\rm exp}{(\beta-\alpha)}(\frac{E}{100\,{\rm keV}})^{\beta},
			 \mbox{if $E > (\alpha-\beta)E_{0}$}
		\end{cases}
		\label{eq:band}
	\end{equation}
	\end{footnotesize}
	Where \emph{A} is the normalization constant, \emph{E} is the energy in unit of keV, $\alpha$ is the low-energy photon spectral index, $\beta$ is the high-energy photon spectral index, and \emph{E$_{0}$} is the break energy in the spectrum.
	The peak energy in the $\nu F_\nu$ spectrum is called $E_{p}$, which is equal to $E_{0}\times(\alpha+2)$.
	In addition, when the count rate of high-energy photons is relatively low, the high-energy spectral index  $\beta$ often cannot be constrained.
	In this case, we consider using a cutoff power-law (CPL) function,
	\begin{equation}
		{ N(E)=A(\frac{E}{100\,{\rm keV}})^{\alpha}{\rm exp}(-\frac{E}{E_c}) },
	\end{equation}
	where \emph{$\alpha$} is the power law photon spectral index, \emph{E$_{c}$} is the break energy in the spectrum,
	and the peak energy $E_{p}$ is equal to $E_{c}\times(2+\alpha)$.
	
	In our joint spectral fitting analysis of Fermi-GBM and Fermi-LAT,
	we use the $pgstat$ statistic \footnote{https://heasarc.gsfc.nasa.gov/xanadu/xspec/manual/\\XSappendixStatistics.html} in Bayesian inference.
	The process used is the same as that described in Section \ref{sec:PDS}.
	The difference is in the model and likelihood function, as well as the additional process of folding the model and the instrumental response file.
	The fitting results of the spectrum are presented in Table \ref{tab:tab1}, and the evolution of the time-resolved spectrum is illustrated in Figure \ref{fig:2} (b).
	Evidently, the spectral evolution does not quite satisfy the hard-to-soft evolution pattern or intensity-tracking pattern \citep{norris1986spectral,golenetskii1983correlation,lu2012comprehensive}.
	Furthermore, the posterior parameter of the spectra analysis of Episode a and Episode b are very similar, which will be further analyzed in Section \ref{sec:lens}.
	
	\subsection{Empirical Correlations and $\rm T_{90}$ Related Distribution}\label{sec:amati}
\begin{figure}
		\centering
		\includegraphics[width=0.48\textwidth]{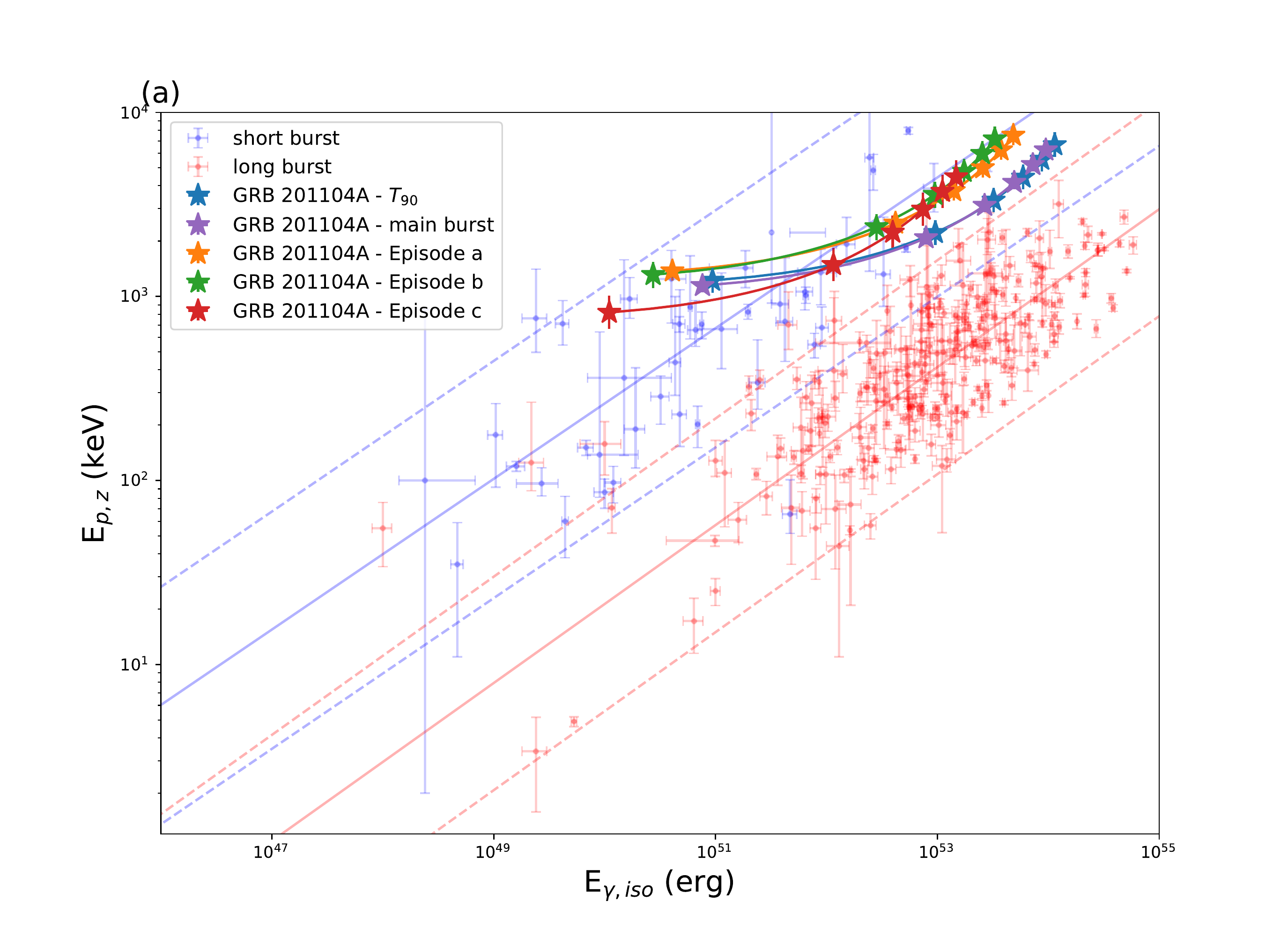}
		\includegraphics[width=0.48\textwidth]{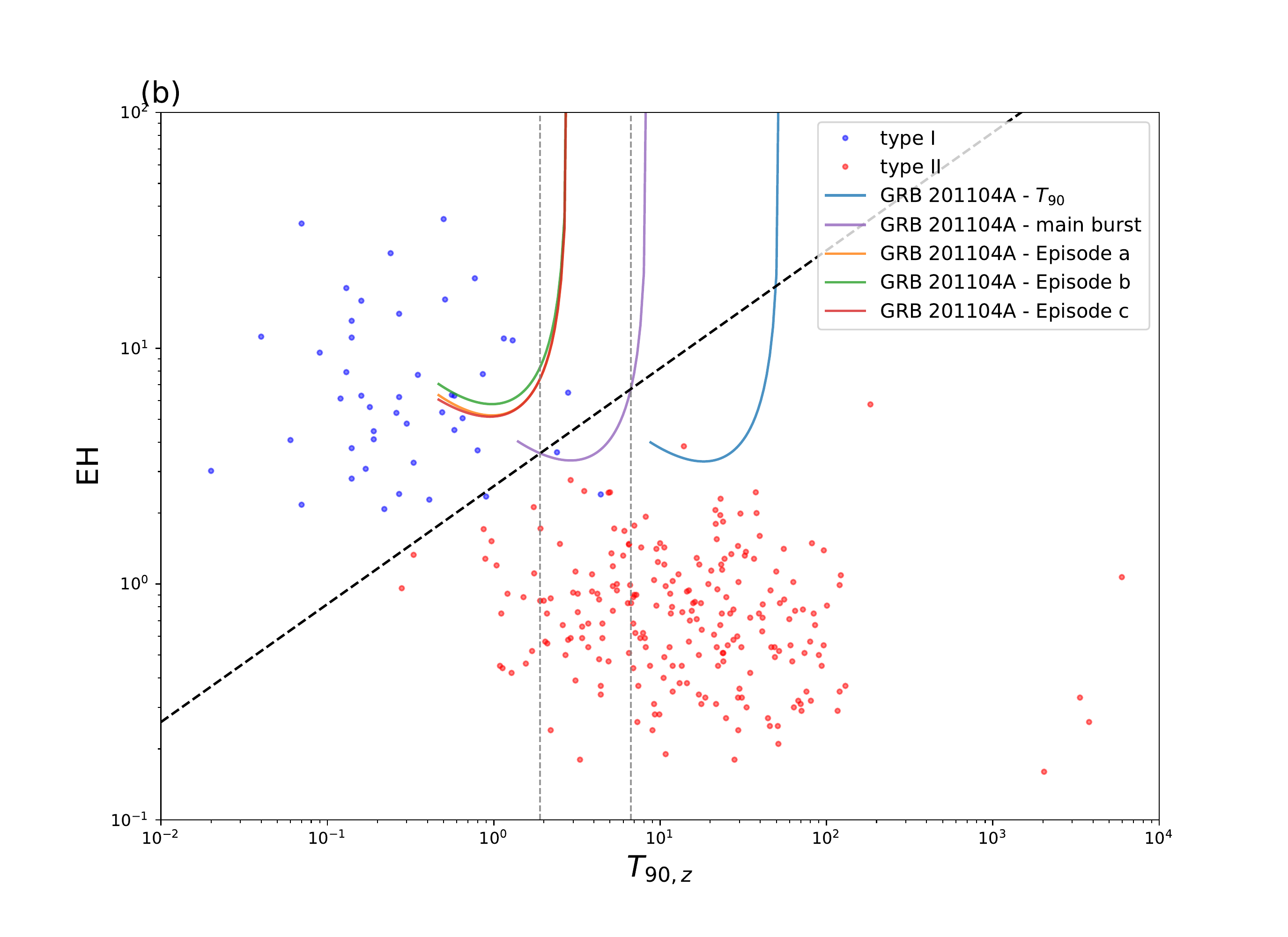}
		\includegraphics[width=0.32\textwidth]{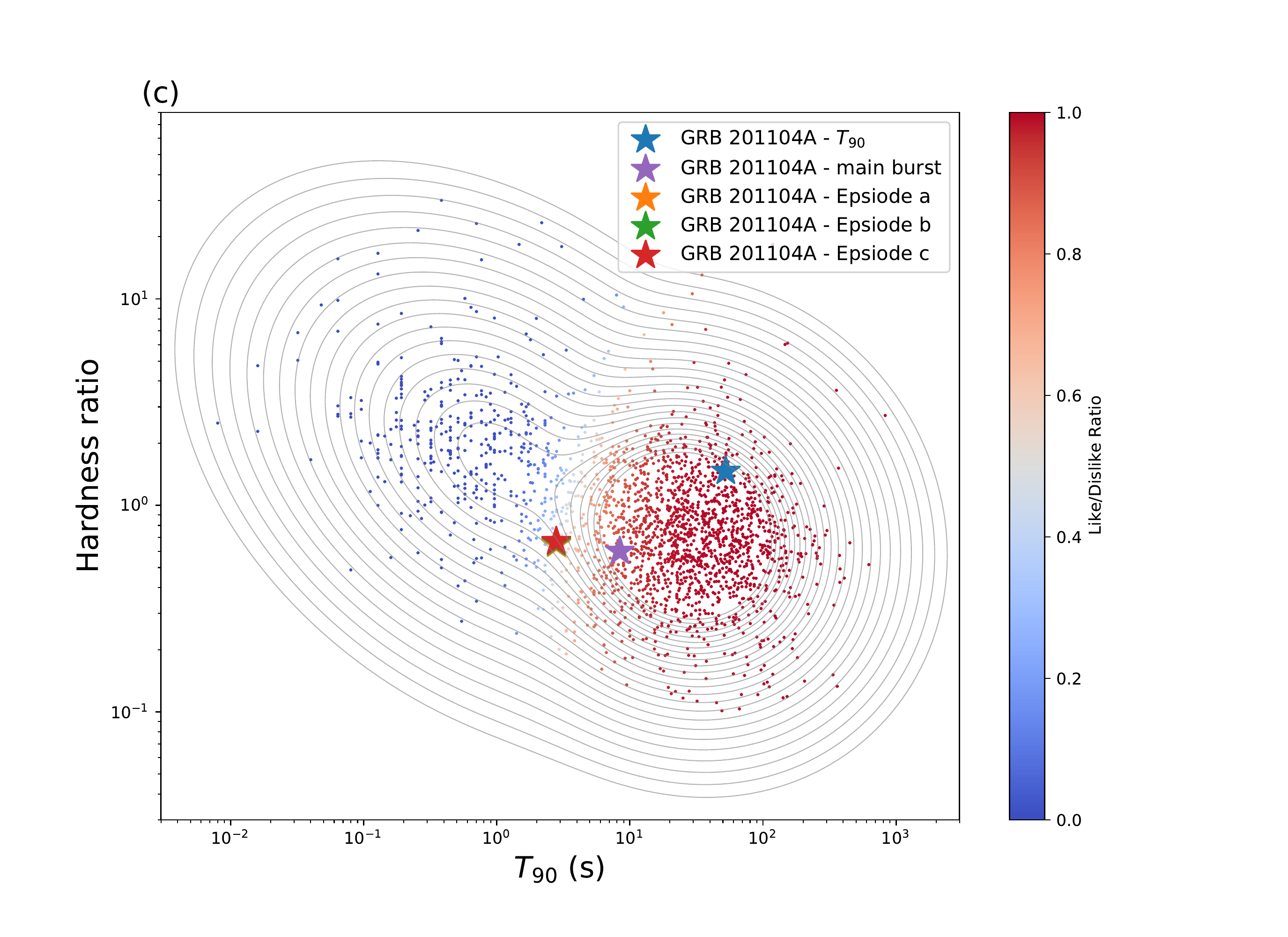}
		\includegraphics[width=0.32\textwidth]{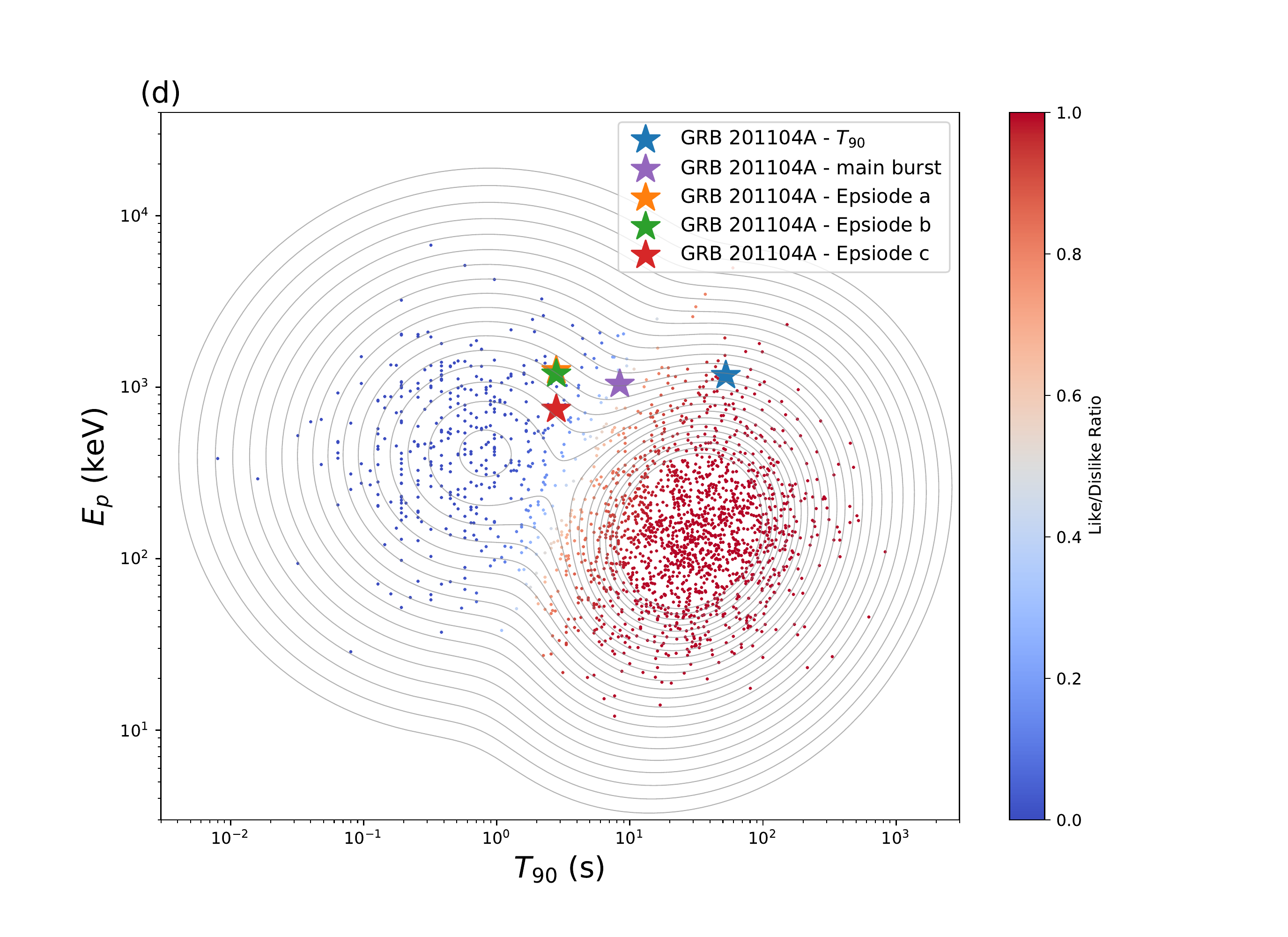}
		\includegraphics[width=0.32\textwidth]{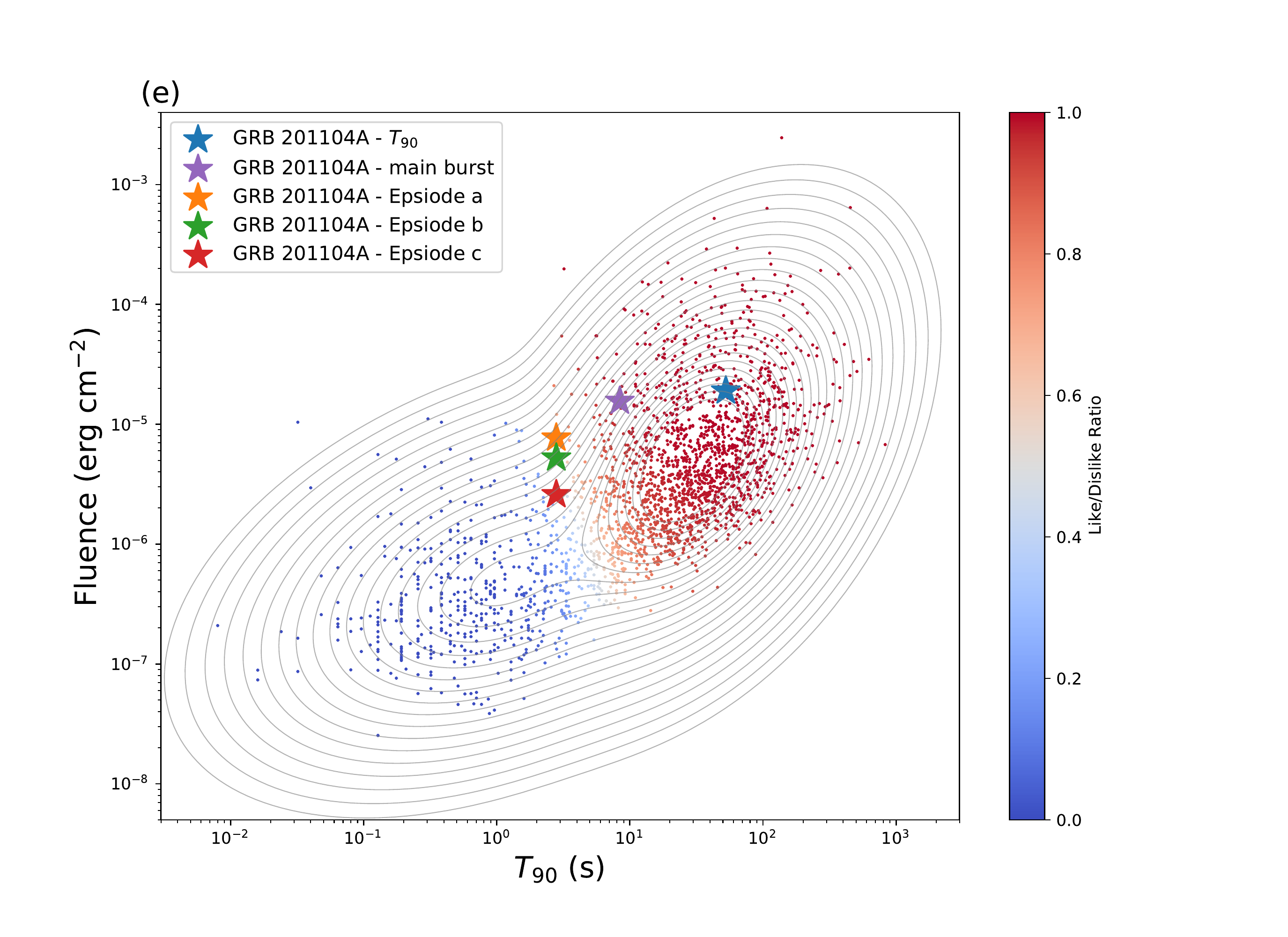}
		\caption{Correlations and statistical distributions.
			(a) is the {redshift} spectral peak energy ($E_{p,z}$) and isotropic equivalent gamma-ray radiation energy ($E_{\gamma,iso}$) correlation diagram.
			The blue and red {solid lines} are the best-fit correlations of the type-I and type-II GRB samples,
			{and the corresponding dashed lines represent the 2$\sigma$ correlation regions respectively.}
			Different colored stars and trajectory indicate each episode in GRB 201104A in different redshift values.
			(b) is the $T_{90,z}$ - $EH$ diagram, and the content represented by each color is the same as in (a).
			{The vertical gray dashed line represents the interval where main butst episode falls within the LGRBs region when $0.26 < z < 3.42$.}
			(c), (d) and {(e)} give the HR, $E_p$ and  {fluence} of each episode respectively, and the comparison with other GRBs in the Fermi catalog.}
		\label{fig:4}
	\end{figure}

	Through the spectral analysis result in Section \ref{sec:spec_ana}, we attempt to classify GRB 201104A using the Amati correlation \citep{amati2002intrinsic,minaev2020p}.
	{Due to the lack of a measured redshift}, we calculated $E_{\gamma, {\rm iso}}$ and $E_{p,z}$ in different redshifts (from 0.1 to 5) with the cosmological parameters of \emph{H$_{0}$} = $\rm 69.6 ~kms^{-1}~Mpc^{-1}$, $\Omega_{\rm m}= 0.29$, and $\Omega_{\rm \Lambda}= 0.71$.
	The star signs of the different colors in Figure \ref{fig:4} (a) display the results obtained with different redshifts for each episode,
	and clearly they all depart from the range of LGRBs (type II).
	
	Additionally, 
	\cite{minaev2020p} proposed a new classification scheme combining the correlation of $E_{\gamma, {\rm iso}}$ and $E_{p,z}$ and the bimodal distribution of $T_{90}$.
	The parameter $EH$ is proposed to characterize the Amati correlation,
	\begin{equation}\label{}
		EH = \dfrac{(E_{p,z}/100{\rm keV})}{(E_{\gamma,iso}/10^{52}{\rm erg})^{0.4}}.
	\end{equation}
	The $T_{90,z}$ - $EH$ trajectories calculated in different redshifts (from 0.01 to 5) for each episode of GRB 201104A are shown in Figure \ref{fig:4} (b).
	
	{
	In the calculation of the above two correlations, we made assumptions about the redshift.
	For the Amati correlation, within our hypothetical redshift range, each episode is an outlier beyond the 2$\sigma$ of the LGRBs classification. 
	In the $T_{90,z}$ - $EH$ correlation, each epsiode is also within the region of the SGRBs.
	But for the main burst, Its trajectory will cross the boundary line between SGRBs (type I) and LGRBs (type II).
	When the redshift is $0.26 < z < 3.42$, it is located in the region of the LGRBs.
	For the $T_{90}$ episode, a higher redshift is required to fall into the classification of SGRBs.
	However, in some current work \citep{paterson2020discovery,o2022deep,hjorth2012optically,vergani2015long,palmerio2019long}, the redshift of the SGRBs category falls between 0.5 and 2.
	Therefore, we propose a possible situation to make the short burst duration longer in Section \ref{sec:lens}.
}

	{We also examine other $T_{90}$-related distributions through the bursts list given by Fermi GBM Burst catalog \citep{gruber2014fermi,bhat2016third,von2014second,von2020fourth}.}
	We collected the $E_p$ {and fluence} of each burst and calculated the hardness ratio (HR),
	which is the ratio of the observed counts in 50–300 keV compared to the counts in the 10–50 keV band \citep{goldstein2017ordinary}.
	In Figure \ref{fig:4} (c), (d) and {(e)}, we plotted the HR, $E_p$ and  {fluence} of each episode for GRB 201104A and other catalog's bursts together,
	and fitting the distribution by using a two-component Gaussian mixture model with {\tt scikit-learn} \citep{scikit-learn}.
	{When the extended emission} is not considered, each episode will gradually approach SGRB (type I) in classification.
	
	\subsection{Spectral Lag}\label{sec:spec_lag}
\begin{figure}
		\centering
		\includegraphics[width=0.7\textwidth]{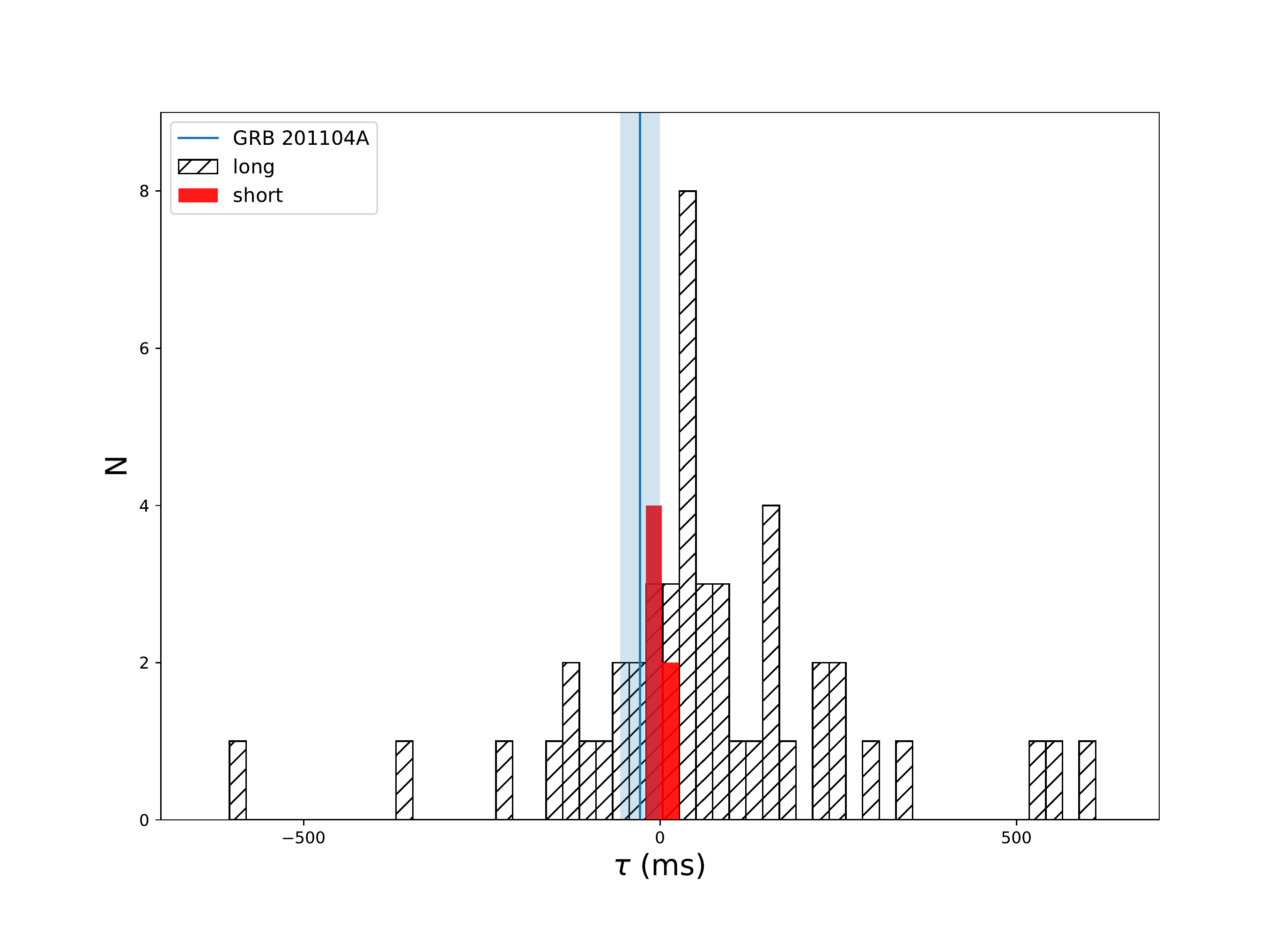}
		\caption{
			{Histograms of spectral lags in the observer frame.
				The black solid line filled histogram and the red filled histogram are the spectral lags of long bursts and short bursts \citep{bernardini2015comparing}, respectively.
				The solid blue line and shaded area are the spectral lag and 1$\sigma$ uncertainty interval of GRB 201104A, respectively.}
		}
		\label{fig:5}
	\end{figure}

	In most GRBs, there is a lag between different energy bands, which is called the spectral lag.
	In general, LGRBs exhibit a relatively significant spectral delay \citep{norris2000connection,gehrels2006new}, {whereas short hard bursts display a negligible spectral lag} \citep{norris2006short}.
	Besides, a fraction of short GRBs even show negative lags \citep{yi2006spectral}.
	A cross-correlation function (CCF) can be used to quantify such an effect since the pulse peaks at different energy bands are delayed.
	{Unlike ACF, CCF is used to compare different time series.}
	The method is widely used to calculate spectral lag \citep{band1997gamma,ukwatta2010spectral}.
	We calculated the CCF function for GRB 201104A in the energy bands {between 100-150 keV and 200-250 keV from $T_0$-0.1 to $T_0$+8.3 s (the main burst episode)},
	and calculated the peak value of CCF after polynomial fitting.
	By using Monte Carlo simulations, we can estimate the uncertainty of lags \citep{ukwatta2010spectral}.
	{The spectral lag of GRB 201104A is ($\tau$ = -28.0 $\pm$ 56.1) ms and is compared with the delays of other long and short GRBs \citep{bernardini2015comparing}, as shown in Figure \ref{fig:5}.}
	{In the observer frame, the mean values of spectral lags for long and short GRBs are ($\tau^L$ = 102.2 $\pm$ 38.1) ms and ($\tau^S$ = -0.73 $\pm$ 7.14) ms \citep{bernardini2015comparing}, respectively.}
	{Although the result of the spectral lag of this burst is closer to the mean value of SGRBs, it cannot be classified accurately.}
	
	\section{Lensing hypothesis}\label{sec:lens}
	To explain the similar spectra of Episode a and Episode b and their long-duration but short-burst characteristics,
	we propose that Episode b is the result of lensing in Episode a and Episode c is a relatively soft and weak extended radiation,
	so no repeats of it have been observed.
	
	In a gravitational lensing system, photons that travel longer distances arrive first,
	because a shorter path means that the light passes through deeper gravitational potential well of the lens, where the time dilation is stronger.
	The source flux is {lower for photons arriving later compared to those which arrive earlier}.
	There will therefore be at least one early pulse followed by a weaker pulse for a lensed GRB.
	The time delay between these two pulses is determined by the mass of the gravitational lens.
	For lensing of a point mass, we have \citep{krauss1991new,narayan1992determination,mao1992gravitational}
	\begin{equation}
		(1+z_\text{l})M_l = \frac{c^3\Delta t}{2G}\left(\frac{r-1}{\sqrt{r}} +\ln r\right)^{-1}.
		\label{eq:mass_redshift}
	\end{equation}
	where $\Delta t$ is the time delay, $r$ is the ratio of the fluxes of the two pulses, and $(1+z_\text{l})M_l$ is the redshifted lens mass.
	With the measured $\Delta t$ and $r$, it is straightforward to calculate the redshifted mass $(1+z_\text{l})M_l$.
	Using Bayesian inference methods of light curve and energy spectrum data, we will estimate the parameters and compare lens and non-lens models.
	
	\subsection{Light-curve Inference}\label{sec:lcinf}
	\cite{paynter2021evidence} developed a Python package called {\tt PyGRB} to create light-curves from either pre-binned data or time-tagged photon-event data.
	We extend the analysis to the Fermi data as well \citep{wang2021grb}.
	Here the same method as in Section \ref{sec:PDS} is used to obtain the posterior distributions of the parameters.
	Bayesian evidence ($\mathcal{Z}$) is calculated for model selection and can be expressed as
	\begin{equation}
		\mathcal{Z} = \int \mathcal{L}(d|\theta) \pi(\theta) d\theta, 
	\end{equation}
	where $\theta$ is the model parameters, and $\pi(\theta)$ is the prior probability.
	For TTE  data from various instruments, the photon counting obeys a Poisson process and
	the likelihood $\ln\mathcal{L}$ for Bayesian inference takes the form of
	\begin{align}
		\ln {\cal L}(\vec{N}|\theta) = & \sum_i
		\ln{\cal L}(N_i|\theta) \\
		= & \sum_i N_i\ln\Big(\delta t_i B +  \delta t_i S(t_i|\theta)\Big) \nonumber\\
		& -  \Big(\delta t_i B +  \delta t_i S(t_i|\theta)\Big) -\log(N_i!),
	\end{align}
	where $N_i$ stands for observed photon count in each time bin, and the model predicted photon count  consists of the background count $\delta t_i B$ and the signal count $\delta t_i S(t_i|\theta)$.
	Note that the differences of $\mathcal{Z}$ among models are important for our purpose. 
	Usually the light curve of a GRB is a pulse shape of fast-rising exponential decay (FRED),
	\begin{equation}
		S(t|\Delta,A,\tau,\xi) = A \exp \left[ - \xi \left(  \frac{t - \Delta}{\tau} + \frac{\tau}{t-\Delta}  \right)   \right],
	\end{equation}
	where $\Delta$ is the start time of pulse, $A$ is the amplitude factor, $\tau$ is the duration parameter of pulse, and $\xi$ is the asymmetry parameter used to adjust the skewness of the pulse. 
	Through different FRED functions, we define different light curve models $S(t_i|\theta)$ to describe whether the pulses are lensed images or not,
	the lensing and null scenarios as
	\begin{align}
		S_\text{lens}(t|\theta_\text{lens}) = &S(t|\Delta,A,\tau,\xi)  \nonumber\\
		&+ r^{-1} \cdot S(t|\Delta+\Delta_t,A,\tau,\xi) + B,
	\end{align}
	\begin{align}
		S_\text{non-lens}(t|\theta_\text{non-lens}) =& S(t|\Delta_1,A_1,\tau_1,\xi_1) \nonumber\\
		&+ S(t|(\Delta_1+\Delta_t,A_2,\tau_2,\xi_2)  + B.
	\end{align}
	For lens model, $r$ is the flux ratio between two pulses (see Equation (\ref{eq:mass_redshift})) and \emph{B} is a constant background parameter.
	The ratio of the $\mathcal{Z}$ for two different models is called as the Bayes factor (BF) and the logarithm of the Bayes factor reads
	\begin{align}
		\ln\text{BF}^\text{lens}_\text{non-lens} = \ln({\cal Z}_\text{lens}) - \ln({\cal Z}_\text{non-lens}) .
	\end{align}
	As a statistically rigorous measure for model selection, 
	if $\ln{\rm(BF)} > 8$ we have the ``strong evidence'' in favor of one hypothesis over the other \citep{thrane2019introduction}.
	
	This method was used to analyze the time series of {Episode a and Episode b}.
	We masked the light curve after Episode b with Poisson background in order to exclude the influence of other time periods.
	In Table \ref{tab:tab2}, the results of Bayesian inference based on the light curves of the NaI and BGO detectors are presented.
	
	\subsection{Pearson Correlation Coefficient}\label{sec:Pearson}
\begin{figure}
		\centering
		\includegraphics[width=0.7\textwidth]{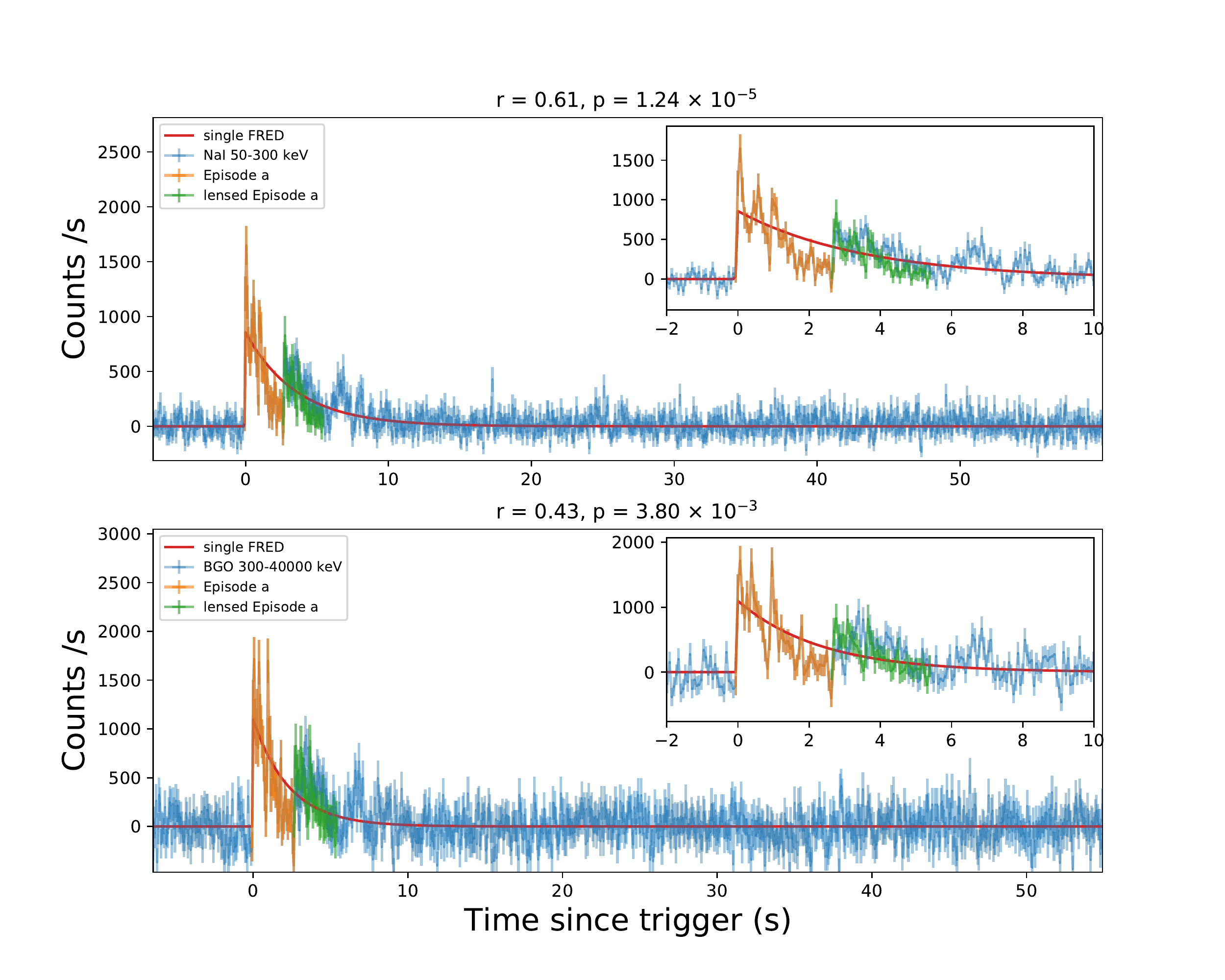}
		\caption{Pearson correlation coefficient.
			{The red line is the best fit result of the light curve with a single FRED model.}
			The green curve (Epsiode b) is the orange curve (Episode a) multiplied by the ratio $r^{-1}$ given by the lens model in Section \ref{sec:lcinf}.
			The upper and lower panel are the two energy bands of NaI (50-300 keV) and BGO (300-40000 keV), respectively.}
		\label{fig:6}
	\end{figure}

	The pulse shape we observe in reality does not correspond to a simple FRED function, and if multiple FRED functions are used to construct the model,
	Bayesian inference will become significantly more difficult.
	Calculating the Pearson correlation coefficient is a simple and effective method of time series analysis.
	Pearson correlation coefficients were calculated for Episode a and Episode b for two different energy bands (see Figure \ref{fig:6}).
	In the NaI detector energy band (50-300 keV), the results are: r = 0.61, p = 1.24 $\times$ $10^{-5}$,
	and in the BGO detector energy band (300-40000 keV), the results are: r = 0.43, p = 3.80 $\times$ $10^{-3}$.
	The results suggest that there is a general correlation between Episodes a and b.
	
	\subsection{Spectral Inference}\label{sec:specinf}
	Under the lensed GRB hypothesis, in addition to requiring the same shape of the light curves, the {similarity} of the spectrum must also be considered.
	The cumulative hardness comparison in different energy bands is a simple but statistically powerful methodology \citep{mukherjee2021hardness}.
	And such a method has been used in some works \citep{wang2021grb,veres2021fermi,lin2021search} as one of the indicators to confirm the lensed GRB.
	
	We propose a procedure that considers spectral fitting with Bayesian inference for lens and non-lens models in order to achieve this goal.
	Typically, the detector response files of GBM contain one or more detector response matrices (DRMs) encoding the energy dispersion and calibration of incoming photons at various energies to recorded energy channels \citep{GbmDataTools}.
	The matrix also encodes the effective area of the detector as a function of energy relative to the source to which the detector is pointing.
	Due to the strong angular dependence of the response (and the degree of angular dependence varies with energy), the effective area can fluctuate significantly.
	Therefore, we should select the DRM that corresponds to the period we are interested in.
	By interpolating the time series DRMs, we can obtain the DRM of the center time of the interaction we are interested in, and generate the corresponding response file.
	It is relatively simple to get the response file of LAT, which is generated by using $gtrepden$ in {\tt Fermitools}.
	Since GRBs are relatively short in duration, so that {the accumulation of LAT background} is negligible.
	To obtain the GBM background file, we use polynomial fitting for the two periods before and after the burst, and then interpolate to obtain the background of the selected time.
	However, although background and instrument responses do not differ significantly in the same GRB event, it is necessary to consider these variations when searching for lensing effects in different GRB events.
	
\begin{figure}
		\centering
		\includegraphics[width=0.48\textwidth]{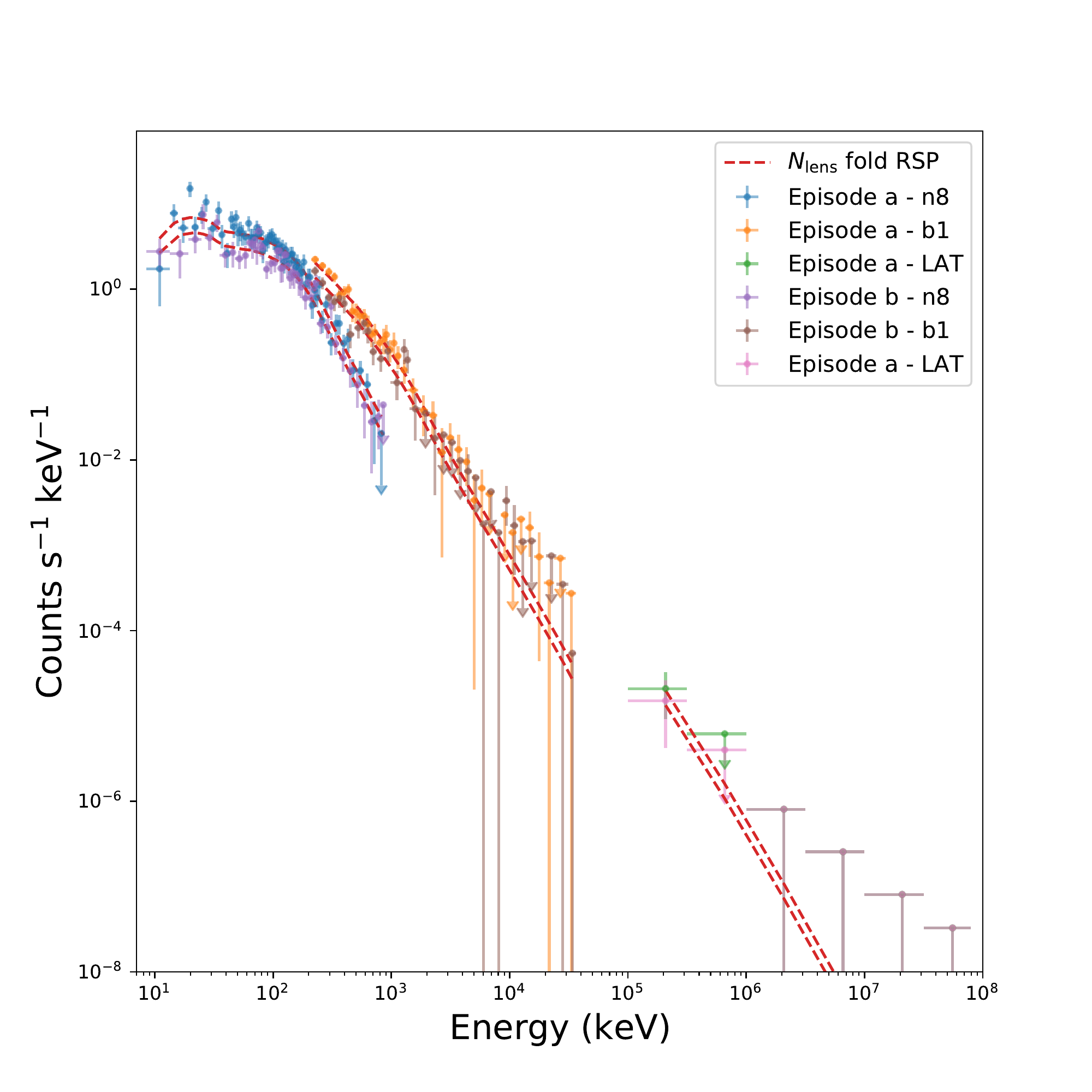}
		\includegraphics[width=0.48\textwidth]{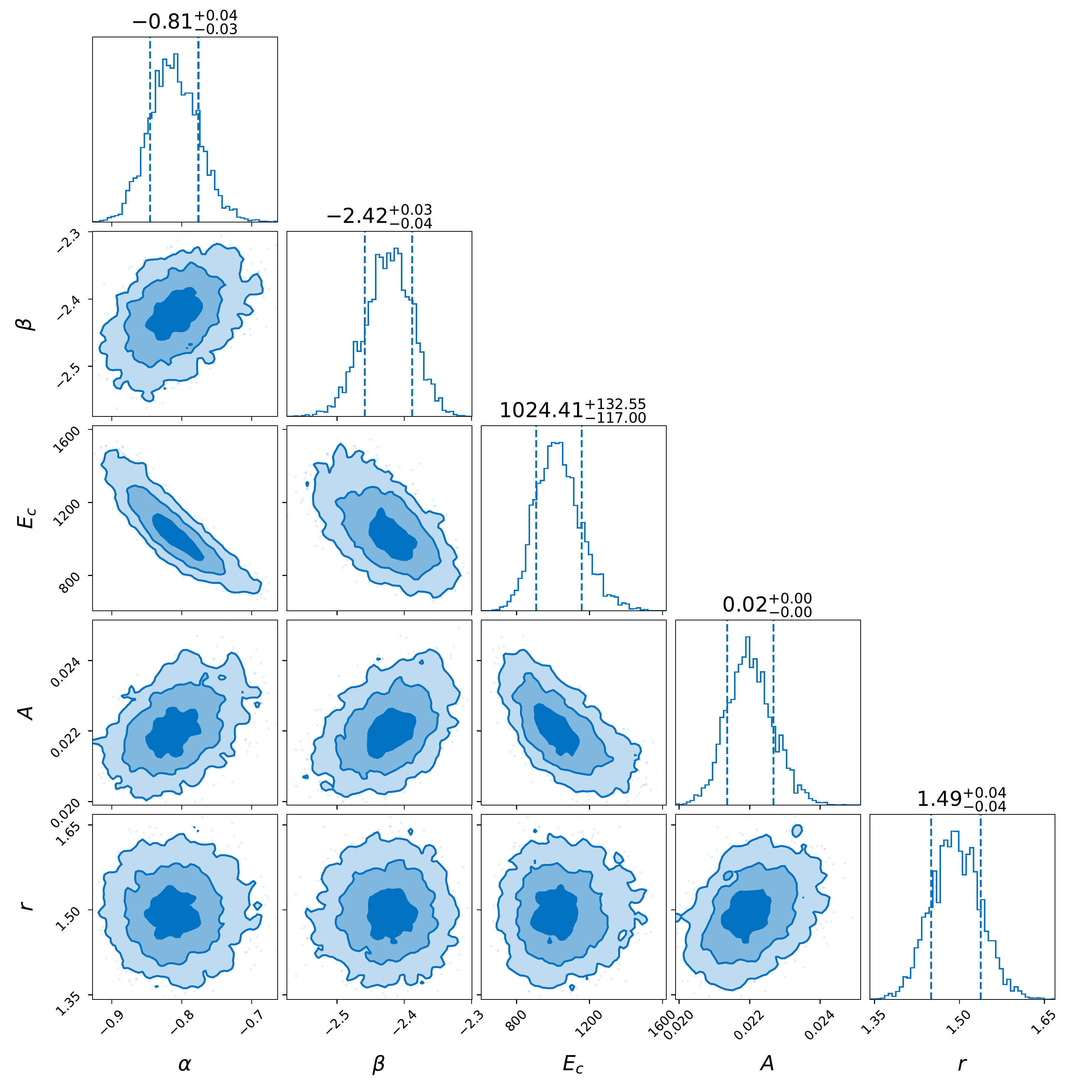}
		\caption{Spectral inference results of lens model. 
			Right: the observed photon count spectra (scatter of different colors) and the photon spectrum (red dashed line) obtained after the lens model folding response file of GBM and LAT. 
			Left: The posterior parameter distribution of the lens model ($N_\text{lens}(E|\theta_\text{lens}) $).}
		\label{fig:7}
	\end{figure}

	The likelihood function used in the spectral inference is the $pgstat$ mentioned in Section \ref{sec:spec_ana}.
	And we use a Band function (Equation \ref{eq:band}) and a ratio parameter $r$ to construct the lens model,
	\begin{align}
		N_\text{lens}(E|\theta_\text{lens}) =& N_\text{Band1}(E|\alpha,\beta,E_c,A) \nonumber  \\
		&+ N_\text{Band2}(E|\alpha,\beta,E_c,A \cdot r^{-1}).
	\end{align}
	The non-lens model is composed of two independent Band functions,
	\begin{align}
		N_\text{non-lens}(E|\theta_\text{non-lens}) = &N_\text{Band1}(E|\alpha_1,\beta_1,E_{c1},A_1) \nonumber  \\
		&+  N_\text{Band2}(E|\alpha_2,\beta_2,E_{c2},A_2).
	\end{align}
	It should be noted that $N_{band1}$ and $N_{band2}$ will fold the response files of Episode a and Episode b respectively.
	We use the above model to fit the spectra {of the two episodes}.
	The method of model comparison is consistent with Section \ref{sec:lcinf}.
	As shown in Table \ref{tab:tab2}, the results of the spectral inference ($\ln{\rm(BF)}= 8.21$) are more inclined to suggest that the spectra of these two time periods differ only in the normalization constant (see Figure \ref{fig:7}).
	{In fact, the spectral evolution of GRBs is very significant and common, and $E_p$-Flux usually has two types of evolution patterns,
		hard-to-soft evolution pattern and intensity tracking pattern \citep{norris1986spectral,golenetskii1983correlation,lu2012comprehensive}, respectively. 
		So for GRB 220411A, the nearly identical spectrum of the first two episodes is peculiar.}
		
	\section{SUMMARY and Discussion}\label{sec:sd}
	In this work, we investigated Fermi's GRBs for possible QPOs and repetitive behavior events, as well as performed a detailed analysis of burst 201104A.
	Our findings are the following:
	\begin{itemize}
		\item Following the current analysis method {in this work}, there is no significant QPO signal greater than 3 $\sigma$ in the light curve of 248 short bursts and 920 long bursts selected.
		However, some GRBs exhibit repetitive behavior, such as GRB 201104A.
		\item GRB 201104A has similar temporal evolutions for both Episodes a and b, as well as the posterior parameters {of the spectral fitting} result.
		According to the Amati correlation diagram, each episode is closer to the group of SGRBs.
		Classifying it as a short burst is also favored in the  $T_{90}$-related distributions when the extended emission component is not taken into account.
		Consequently, GRB 201104A is a long-duration burst with characteristics of SGRBs.
		\item The Bayesian inferences of the light curve do not fully support the lens model. 
		Nevertheless, the spectral inference result supports the lens model, at least showing that the spectra of these two episodes are very consistent.
	\end{itemize}
	
	A long-duration burst with the characteristics of a short burst can be explained very naturally by the gravitational-lensing scenario, although this is a very coincidental circumstance.
	As well, it is not impossible that this event may be just a very special occurrence that comprises at least two intrinsically similar episodes.
	In other words, there is a repetition mechanism in GRB, like the central engine with memory that \cite{zhang2016central} once proposed.
	This type of burst has many features characteristic of a SGRB,
	but unless the Li-Paczynski macronova (also known as the kilonova) can be detected \citep{li1998transient},
	this will undoubtedly be the strongest evidence for a SGRB of compact-binary origin, {such as GRB 060614 \citep{gehrels2006new,yang2015possible,jin2015light} and GRB 211211A \citep{rastinejad2022kilonova,gompertz2022minute,yang2022peculiarly}.}
	
	Furthermore, our proposed procedure is essentially consistent with the method proposed by \cite{mukherjee2021hardness} for comparing the cumulative hardness in different energy bands.
	In this procedure, we use the Bayes factor to determine the consistency of the spectrum.
	Over time series analysis, the advantage is that low count rate instruments can be considered in Bayesian inference, such as the spectra of high-energy photons detected by Fermi-LAT.
	For certifying different GRB events as gravitational lensing events, this procedure could take into account the effects of instrumental responses as well as background.
	There is currently a problem with this procedure, which is that the episodes used for spectral inference must be selected in advance.
	Therefore, this procedure will only give a posterior distribution of the ratio $r$ without a time delay caused by gravitational lensing.
	We can set the time delay and time window as free parameters to solve this problem.
	Since each model calculation must take into account the change in the spectral fitting file, the calculation time will greatly increase.
	Our future work will optimize this step in order to search for gravitationally lensed GRBs.
	Currently, the most complete procedure is to use both time series and spectrum data to perform Bayesian inference for lens and non-lens models.
	
	\section*{Acknowledgments}
	We appreciate the anonymous referee for their helpful suggestions. 
	We thank S. Covino and Yi-Zhong Fan for their important help in this work.
	We appreciate Zi-Min Zhou for his helpful suggestions.
	We acknowledge the use of the Fermi archive's public data.
	This work is supported by NSFC under grant No. 11921003.
	
	\software{\texttt{Matplotlib} \citep{Hunter:2007}, \texttt{Numpy} \citep{harris2020array}, \texttt{scikit-learn} \citep{scikit-learn},
	\texttt{bilby} \citep{ashton2019bilby}, \texttt{GBM Data Tools} \citep{GbmDataTools}, \texttt{Fermitools}}
	
	\bibliography{bibtex}
	\clearpage
	\setlength{\tabcolsep}{2mm}{}
	\begin{deluxetable}{cccccccccccccccc}\tiny
		\tabletypesize{\tiny}
		\tablecaption{Spectral fitting results}
		\tablehead{ \colhead{}& \colhead{Detector}&\colhead{} &\colhead{Band or CPL Function} &\colhead{} &\colhead{Fluence (10-1000 keV)} \\
			\colhead{Time Interval (s)} &  \colhead{} &\colhead{$\alpha$}  &\colhead{$\beta$}&\colhead{$E_{\rm p}$ (keV)} &\colhead{$10^7$ erg $\rm cm^{-2}$}}
		\startdata
		time-integrated spectra   \\
		$T_{90}$  {[-0.10 - 52.50]} & GBM+LAT & -1.03$_{-0.04}^{+0.05}$  & -2.27$_{-0.03}^{+0.03}$ & 1111.37$_{-157.15}^{+189.42}$ & 190.21$_{-4.90}^{+5.11}$\\
		main burst  {[-0.10 - 8.30]} & GBM+LAT & -0.78$_{-0.03}^{+0.04}$  & -2.40$_{-0.03}^{+0.03}$ & 1041.21$_{-85.26}^{+91.65}$ & 156.10$_{-2.40}^{+2.37}$\\ 
		\hline
		\object time-resolved spectra   \\
		Episode a {[-0.10 - 2.70]} & GBM+LAT & -0.84$_{-0.04}^{+0.05}$  & -2.42$_{-0.05}^{+0.05}$ & 1250.33$_{-141.46}^{+168.67}$ & 76.99$_{-1.47}^{+1.58}$\\
		{  [-0.10 - 0.12]} & GBM &-1.03$_{-0.06}^{+0.05}$  & ... & 1897.09$_{-329.76}^{+398.86}$ & 5.46$_{-0.25}^{+0.24}$\\
		{  [0.12 - 1.15]} & GBM &-0.89$_{-0.03}^{+0.03}$  & ... & 1996.96$_{-190.91}^{+209.56}$ & 22.29$_{-0.55}^{+0.57}$\\
		{  [1.15 - 1.60]} & GBM &-0.82$_{-0.11}^{+0.10}$  & ... & 1139.32$_{-245.12}^{+353.71}$ & 5.25$_{-0.30}^{+0.30}$\\
		{  [1.60 - 2.70]} & GBM &-0.80$_{-0.20}^{+0.19}$  & ... & 724.01$_{-193.49}^{+515.03}$ & 4.92$_{-0.38}^{+0.40}$\\
		Episode b {[2.70 - 5.50]} & GBM+LAT & -0.78$_{-0.06}^{+0.07}$  & -2.44$_{-0.06}^{+0.06}$ & 1194.15$_{-183.33}^{+205.28}$  & 52.47$_{-1.36}^{+1.47}$\\
		{  [2.70 - 3.80]} & GBM &-0.95$_{-0.05}^{+0.05}$  & ... & 2189.62$_{-386.93}^{+443.95}$ & 13.11$_{-0.51}^{+0.47}$\\
		{  [3.80 - 5.50]} & GBM &-0.74$_{-0.09}^{+0.09}$  & ... & 1188.96$_{-185.32}^{+248.81}$ & 13.08$_{-0.55}^{+0.55}$\\
		Episode c {[5.50 - 8.30]} & GBM+LAT & -0.74$_{-0.11}^{+0.13}$  & -2.38$_{-0.08}^{+0.07}$ & 744.10$_{-138.81}^{+174.84}$ & 25.83$_{-1.28}^{+1.29}$\\
		{  [5.50 - 6.80]} & GBM &-0.88$_{-0.13}^{+0.12}$  & ... & 1063.12$_{-267.90}^{+423.17}$ & 6.76$_{-0.45}^{+0.43}$\\
		{  [6.80 - 8.30]} & GBM &-0.79$_{-0.15}^{+0.14}$  & ... & 744.63$_{-160.18}^{+291.04}$ & 6.24$_{-0.45}^{+0.48}$\\
		\enddata
		\tablecomments{When beta is ``..." indicates that the CPL function is used for fitting.}
		\label{tab:tab1}
	\end{deluxetable}
	\setlength{\tabcolsep}{2mm}{}
	\begin{deluxetable}{cccccccccccccccc}\tiny
		\label{tab:tab2}
		\tabletypesize{\tiny}
		\tablecaption{Bayesian inference result}
		\tablehead{ \colhead{dectors}& \colhead{$\ln{\cal Z}_\text{lens}$}&\colhead{$\ln{\cal Z}_\text{non-lens}$} &\colhead{$\ln$(BF)} &\colhead{favourite model} }
		\startdata
		Light curve inference\\
		GBM-n8 (50-300 keV) & -1114.03 & -1110.59  & -3.44 &  $S_\text{non-lens}$ \\
		GBM-b1 (300-40000 keV) & -1284.27 & -1285.83 & 1.56 &  $S_\text{lens}$   \\ 
		\hline
		Spectral inference\\
		GBM (n8, b1) + LAT & -519.45  & -527.66 &  8.21 & $N_\text{lens}$  \\	
		\enddata
	\end{deluxetable}
\end{document}